\documentstyle[preprint,aps]{revtex}
\tightenlines
\begin{document}
\draft
\title
{\bf Comments on some recent papers on the hyperspherical approach
in few--body systems}
\author{V.D. Efros}
\address{
Russian Research Centre "Kurchatov Institute", Kurchatov Square 1,
123182 Moscow, Russia}
\date{\today}
\maketitle
\begin{abstract}
Recent papers by Kievsky, Viviani, and Rosati are commented. 
It is pointed out
that the techniques for treatment few--body reactions and 
bound states 
in the framework of the hyperspherical approach 
presented in those papers as original ones were in fact  
known and have been applied previously.
\end{abstract}
 
\vspace{5mm}

The purpose of these notes is to demonstrate the fact of 
misleading presentation 
in the recent papers by Kievsky, Viviani, and Rosati 
(Refs.\cite{KVR95,KVR97,K97,KMVR97}). Namely, some aspects 
of the hyperspherical harmonic (HH)
method in few--body systems are presented in those papers as 
original ones, though,
in reality, they have been developed in the literature previously.
And the authors were aware of the latter (which can be proved).
Just this, and not a
mere point of priority, deserves 
a reaction. Therefore 
I feel obliged to discuss this issue in public.

Below the technique for treatment reactions in the framework of the
(HH) approach is discussed. In Appendix the procedure for reduction 
of the number of HH entering calculations is considered.

It is 
not my aim 
to discuss here the usefulness of the large--scale calculations 
performed by the above mentioned group. 

\section*{Treatment of reactions in the framework of the HH approach}

The paper \cite{KVR95} by Kievsky, Viviani, and Rosati starts with:

"In a recent paper \cite{KVR94} (by the same authors, {\em V.D.E.}) 
a variational technique to calculate
scattering states below the deutron breakup threshold 
has been developed
for the three--nucleon $N-d$ system. In this approach, the 
wave function 
of the system is written as a sum of two terms
\begin{equation} 
\Psi=\Psi_C+\Psi_A.
\end{equation}
The first term $\Psi_C$ is responsible for the description of the 
system when the three nucleons are close to each other. It is 
decomposed in channels labeled by the angular--spin--isospin 
quantum numbers and the corresponding two--dimensional spatial 
amplitudes are expanded in terms 
of the pair correlated hyperspherical harmonic (PHH) basis 
\cite{KVR93}. The
second term $\Psi_A$ is a solution of the 
Schr\"odinger equation in
the asymptotic region where the incident nucleon 
and the deutron are
well apart."
 
This statement, repeated in subsequent papers by Kievsky
et al. \cite{KVR97,K97}, is misleading since in 
reality the quite similar technique
was used before in the literature. 
The above extract is to be compared to that from 
the introductory section of 
Ref. \cite{ZPE68}:

"\ldots In the present work the $K$--harmonics 
method is generalized 
to the scattering problem for the three--nucleon 
system. However, in
that case in no approximation in the expansion of 
the wave function can one
ignore harmonics with large $K$. This is due to 
the fact that that part of
the wave function which corresponds to the 
asymptotics of two--fragment
channels (free motion of the nucleon with 
respect to the deutron) requires for
its description a large number of harmonics.

It is natural to write $\Psi$ in the form
\begin{equation}
\Psi=\Phi+X,
\end{equation}
where $\Phi$ describes the two--fragment of $\Psi$, and 
to search for
an expansion in $K$ harmonics of the part $X$ only. 
It then turns out that it 
is possible to restrict oneself to a small number of 
terms in such an expansion.

The function $\Phi$ is known accurate to within the 
sought for amplitudes 
$f_L$\ldots"

Taking 
into account that "$K$-harmonics" 
mean the HH basis, the main difference between the 
two extracts lies merely in the notation in 
Eqs. (1) and (2). A minor
difference is that in the latter case the pure HH 
basis is discussed while
in the former one the correlated HH basis is 
mentioned. Obviously,  
the correlation factors are included into the wave 
function exactly in the same way as for bound states, and 
this has nothing 
to do with the specific character of the continuum 
spectrum problem. 
(In fact, the
correlated HH basis was  considered first in 1972 in 
Ref. \cite{FE72} in  
the  
bound state 
problem. It has been shown there that the inclusion 
of the correlations
leads to a great increase in the convergence rate. In 
the commented 
papers the 
correlations are applied in a modified form.)

The convenience of the HH expansion for the 
description of the three--body
breakup reaction has been recognized for the first 
time in Ref. \cite{Delv59}.
However, the presence of the two--fragment 
reaction channels led to
difficulties. To solve the problem, in the subsequent paper  
\cite{Delv61} 
the author  attempted to combine the HH approach
and the resonating group approach. This, however, led to 
integro-differential equations being very involved even 
when only the lowest HH term is retained. In 
the next paper \cite{Delv62} 
the author renounced of the treatment of the 
two--body channels at all
simply disregarding them. 
But that  
led to unrealistic results for the three--body 
photodisintegration cross section.
(See Ref.  \cite{Lev79} on possible approximate 
interpretation of such type calculations.)
One more attempt to generalize the HH approach 
to the scattering problem was undertaken 
in Ref. \cite{Zic67} independently
of Ref. \cite{ZPE68} but it was also not successful enough. 
The difficulty
has been removed in the above mentioned paper \cite{ZPE68} 
where the procedure 
leading to a simple set of simultaneous differential 
equations for the 
coefficients of the HH expansion
and algebraic equations for the scattering or 
reaction amplitudes has been
given. It corresponds to the first order 
Hulth\'en--Kohn type calculation. 
The second order variational Hulth\'en--Kohn 
corrections were calculated 
in Refs. \cite{PPFE71,EZ71,Ray73}. Among the 
other papers on the subject, 
we mention Ref. \cite{VZ81} 
where the photodisintegration of $^3$H was studied 
using the
modified version \cite{EZ71} of the procedure, and 
the calculations above
the breakup threshold were performed with a 
noncentral realistic NN
potential of those days. 

Claiming the developement of a variational technique 
to calculate
scattering states, Kievsky, Viviani, and Rosati have never 
mentioned 
the papers \cite{ZPE68,PPFE71,EZ71,Ray73,VZ81}. 
While the studies performed in those papers required less 
intensive calculations 
than those performed by Kievsky et al., the whole 
procedure itself for 
treatment continuum states 
used by the latter authors in Refs. \cite{KVR94,KVR95} 
was essentially 
the same as in
Refs. \cite{ZPE68,PPFE71}. I did not 
find any new point concerning 
that procedure 
in \cite{KVR94,KVR95}, and hence 
there is no possibility to
understand what  
the variational technique to calculate
scattering states "has been developed" there.

I have recently got to know from my Russian 
colleagues\footnote{I could have listed
the names.} that they in fact discussed the considered 
approach with Viviani at the 
Uzhgorod Few--Body Conference
in 1990 and provided him with the correspoding references, 
in particular. In addition, in 1991 they visited Pisa and 
delivered a University seminar on HH three--body 
calculations of
continuum and bound states.
Kievsky, Viviani and Rosati were present 
at the seminar, and 
had further discussions 
with them on
the procedure in question during that visit. Therefore, 
concerning the 
presentation in
Refs. \cite{KVR95,KVR97,K97}, there is no question whether 
the "Pisa group" was aware
of the preceding literature on the subject while performing 
their  calculations. 

\section*{Appendix.
An algorithm for reduction of the number of  HH basis states }

A quite similar comment applies to the paper
\cite{KMVR97} by Kievsky et al. 
In that paper, the authors have managed to solve accurately 
the bound--state three nucleon 
 problem for the Argonne  $NN$ interaction 
due to selecting in 
the calculation  certain subsystems from complete sets of 
HH at given $K$ values.
The authors discuss their selection comparing it with other 
"various selections of the HH functions" in the literature 
but do not mention
that the selection prescription they are just using 
was known before.  
This prescription for selecting (or adding consecutively) 
the subsystems 
has in fact been proposed in Ref. \cite{2}.

Earlier than in Ref. \cite{KMVR97} by Kievsky et al. 
this prescription 
has been 
used in Refs. \cite{3,4,5} and 
\cite{6} 
in the three-- and four--nucleon calculations, respectively. 
In Ref. \cite{KMVR97} 
the first of these papers is mentioned. It is stated 
there that Ref. \cite{3}, among others,  "...revealed a 
slow rate of convergence 
and provided satisfactory results only for soft NN interaction 
models. It turned out that the main difficulty to be 
overcome arises from 
the large degeneracy of the HH basis." This presentation 
is misleading 
since the authors do not say that 
in
Ref. \cite{3}  the prescription for the selection of important 
contributions has already been used and the difficulty with the 
slow rate of convergence 
{\em has been} overcome with its help. Moreover, the 
selection used by the authors
of the commented paper 
is in fact 
{\em identical} to that used in Refs. \cite{2,3,4,5,6}.

This can be clarified by mentioning the following point.
Despite the terminological  differences with the
previous papers, such as use of the term "Faddeev decomposition"
by Kievsky et al., the ansatz for the wave function they use
is in fact identical to that used 
previously\footnote{A slight and only  
apparent 
difference is that in \cite{2,3,4,5,6} a 
"symmetrization" of the 
 spatial channel basis functions is performed,  
and then the functions 
with proper symmetries obtained are coupled to 
the spin--isospin 
basis functions 
with conjugate symmetries to obtain the totally antisymmetric 
basis functions. While in Ref. \cite{KMVR97} products of the 
spatial and spin--isospin 
channel basis functions are symmetrized directly. 
The resulting basis 
functions may be grouped
in pairs spanning two--dimensional subspaces that are obviously 
just the same for the both 
ways of the antisymmetrization. In particular, this 
applies to "important" 
basis functions retained
in \cite{KMVR97} which are thus completely equivalent 
to those defined in Ref. \cite{2}.
The choice of different $K_{max}$ values for HH components 
of different types was done e.g.
in Refs \cite{4,6}.}. The dynamical
equations are also the same. Apart from the inserted 
correlation factors, the same applies to other papers by
Kievsky et al.

(In reality, in the above mentioned earlier papers
by the Russian group 
physically accurate bound state A=3 and 4
results had been obtained with the HH approach for 
softer core
realistic $NN$ interactions. For A=3 they compare well with 
more accurate Faddeev results obtained by that time \cite{Friar91}.
While in Ref. \cite{KMVR97} by Kievsky et al. the 
authors solved the problem 
with the Argonne potential (with the accuracy that far 
exceeds a physically meaningful one)  
due solely to substantial increase in the number of the 
retained HH within the same known selection prescription.)

\end{document}